\documentclass[english]{article}
\usepackage[T1]{fontenc}
\usepackage[latin9]{inputenc}
\usepackage{geometry}
\usepackage{color}
\usepackage{babel}
\usepackage{float}
\usepackage{amsmath}
\usepackage{amsthm}
\usepackage{graphicx}
\usepackage{setspace}
\usepackage[unicode=true]
 {hyperref}
\begin{document}
\begin{center}
\textbf{\Large{}Dynamic Multi-Factor Bid-Offer Adjustment Model }
\par\end{center}{\Large \par}

\begin{center}
\textbf{\large{}A Feedback Mechanism for Dealers (Market Makers) to
Deal (Grapple) with the Uncertainty Principle of the Social Sciences }
\par\end{center}{\large \par}

\begin{center}
\textbf{Ravi Kashyap}
\par\end{center}

\begin{center}
\textbf{City University of Hong Kong / Gain Knowledge Group }
\par\end{center}

\begin{center}
\textbf{Originally Created, October 2008; Revised, October 2013}
\par\end{center}

\begin{center}
\textbf{\textcolor{blue}{\href{http://www.iijournals.com/doi/abs/10.3905/jot.2014.9.3.042}{Edited Version: Kashyap, R. (2014). Dynamic Multi-Factor Bid-Offer Adjustment Model. The Journal of Trading, 9(3), 42-55.}}}\tableofcontents{}
\par\end{center}

\section{\quad Abstract}

The objective is to come up with a model that alters the Bid-Offer,
currently quoted by market makers, that varies with the market and
trading conditions. The dynamic nature of financial markets and trading,
as the rest of social sciences, where changes can be observed and
decisions can be taken by participants to influence the system, means
that our model has to be adaptive and include a feedback loop that
alters the bid offer adjustment based on the modifications we are
seeing in the market and trading conditions, without a significant
time delay. We will build a sample model that incorporates such a
feedback mechanism and also makes it possible to check the efficacy
of the changes to the quotes being made, by gauging the impact on
the Profits. The market conditions here refer to factors that are
beyond the direct control of the market maker and this information
is usually available publicly to other participants. Trading conditions
refer to factors that can be influenced by the market maker and are
dependent on the trading book being managed and will be privy only
to the market maker and will be mostly confidential to others. The
factors we use to adjust the spread are the price volatility, which
is publicly observable; and trade count and volume, which are generally
only known to the market maker, in various instruments over different
historical durations in time. The contributions of each of the factors
to the bid-offer adjustment are computed separately and then consolidated
to produce a very adaptive bid-offer quotation. The ensuing discussion
considers the calculations for each factor separately and the consolidation
in detail. 

Any model that automatically updates the quotes is more suited for
instruments that have a high number of transactions within short intervals,
making it hard for traders to manually monitor and adjust the spread;
though this is by no means a stringent requirement. We can use similar
models for illiquid instruments as well and use the quotations provided
by the model as a baseline for further human refinement. 

We have chosen currency markets to build the sample model since they
are extremely liquid, Over the Counter (OTC), and hence trading in
them is not as transparent as other financial instruments like equities.
The nature of currency trading implies that we do not have any idea
on the actual volumes traded and the number of trades. We simulate
the number of trades and the average size of trades from a log normal
distribution. The parameters of the log normal distributions are chosen
such that the total volume in a certain interval matches the volume
publicly mentioned by currency trading firms. This methodology can
be easily extended to other financial instruments and possibly to
any product with an ability to make electronic price quotations or
even be used to periodically perform manual price updates on products
that are traded non-electronically. 

Thankfully, we are not at a stage where Starbucks will sell coffee
using such an algorithm, since it can possibly lead to certain times
of the day when it can be cheaper to have a cup of coffee and as people
become wary of this, there can be changes to their buying habits,
with the outcome that the time for getting a bargain can be constantly
changing; making the joys of sipping coffee, a serious decision making
affair.

\section{Motivation for Multi-Factor Bid-Offer Models}

At the outset, let us look at some fundamentals that govern all financial
instruments and then delve into the nuances which apply to instruments
that are more amenable to adaptive bid-offer models. It is also worthwhile
to mention here that for most assertions made below, numerous counter
examples and alternate hypothesis can be produced. These are strictly
attempts at tracing the essentials rather than getting bogged down
with a specific instance. However, building a model for empirical
usage requires forming a conceptual framework based on the more common
observations, yet being highly attuned to any specifics that can stray
from the usual. Also, for the sake of brevity, a number of finer points
have been omitted and certain simplifying assumptions have been made. 

The various financial instruments that exist today can be broadly
viewed upon as vehicles for providing credit and a storage for wealth,
for both individuals and institutions alike. The different instruments,
both in terms of their nomenclature and their properties, then merely
become manifestations of which and how many parties are involved in
a transaction and the contractual circumstances or the legal clauses
that govern the transaction. 

Despite the several advances in the social sciences and in particular
economic and financial theory, \textbf{\textit{we have yet to discover
an objective measuring stick of value, a so called, True Value Theory}}.
While some would compare the search for such a theory, to the medieval
alchemists\textquoteright{} obsession with turning everything into
gold, for our present purposes, the lack of such an objective measure
means that the difference in value as assessed by different participants
can effect a transfer of wealth. This forms the core principle that
governs all commerce that is not for immediate consumption in general,
and also applies specifically to all investment related traffic which
forms a great portion of the financial services industry and hence
the mainstay of market making. 

Although, some of this is true for consumption assets; because \textbf{\textit{the
consumption ability of individuals and organizations is limited and
their investment ability is not}}, the lack of an objective measure
of value affects investment assets in a greater way and hence investment
assets and related transactions form a much greater proportion of
the financial services industry. Consumption assets do not get bought
and sold, to an inordinate extent, due to fluctuating prices, whereas
investment assets will. Hull {[}1999{]} has a description of consumption
and investment assets, specific to the price determination of futures
and forwards. The price effect on consumptions assets affects the
quantity bought and consumed, whilst with investment assets, the cyclical
linkage between vacillating prices and increasing number of transactions
becomes more apparent. 

\textbf{\textit{Another distinguishing feature of investment assets
is the existence or the open visibility of bid and ask prices.}} Any
market maker for investment assets quotes two prices, one at which
he is willing to buy and one at which he is willing to sell. Consumption
assets either lack such an outright two sided quote; or it is hard
to painlessly infer viewable buy and sell prices, since it involves
some conversion from a more basic form of the product into the final
commodity being presented to consumers. Examples for consumption assets
are a mug of hot coffee, that requires a certain amount of processing
from other rudimentary materials before it can be consumed; or a pack
of raw almonds which is almost fit for eating. Coffee shops that sell
coffee do not quote a price at which they buy ready drinkable coffee;
the price at which a merchant will buy almonds is not readily transparent.
Gold is an example of both, a consumption and an investment asset.
A jewellery store will sell gold and objects made of gold; but it
will also buy gold reflecting its combined consumption and investment
trait. This leaves us with financial securities like stocks and bonds
that are purely investment assets. 

A number of disparate ingredients contribute to this price effect;
like how soon the product expires and the frequent use of technology
to facilitate a marketplace. EBay is an example of a business where
certain consumption goods are being bought and sold. This can happen
even if goods are only being sold, through the increased application
of technology in the sales process. While not implying that the use
of technology is bad, technology, or almost anything else, can be
put to use that is bad. Thankfully, we are not at a stage where Starbucks
will buy and sell coffee, since it can possibly lead to certain times
of the day when it can be cheaper to have a cup of coffee and as people
become wary of this, there can be changes to their buying habits,
with the outcome that the time for getting a bargain can be constantly
changing; making the joys of sipping coffee, a serious decision making
affair. Even though this is an extreme example, we will overlook some
of these diverse influences for now, since our attempt is to exemplify
the principal differences between the varieties of financial transactions
and the underlying types of assets that drive these deals. 

This lack of an objective measure of value, (henceforth, value will
be synonymously referred to as the price of an instrument), makes
prices react at varying degrees and at varying speeds to the pull
of different macro and micro factors. The greater the level of prevalence
of a particular instrument (or even a particular facet of an instrument)
the more easily it is affected by macro factors. This also means that
policies are enforced by centralized institutions, (either directly
by the government or by institutions acting under the directive of
a single government or a coalition of governments), to regulate the
impact of various factors on such popular instruments. Examples for
this would be interest rate dependent instruments, which are extremely
sensitive to rates set by central banks since even governments issue
such instruments; dividends paid by equity instruments which are clearly
more sensitive to the explicit taxation laws that govern dividends
than to the level of interest rates; and commodities like oil, which
are absolutely critical for the smooth functioning of any modern society
and hence governments intervene directly to build up supplies and
attempt to control the price. See Tuckman {[}1995{]} for interest
rate instruments; Bodie, Kane and Marcus {[}2002{]} for equity instruments. 

Lastly, it is important that we lay down some basics regarding the
efficiency of markets and the equilibrium of prices. Surely, a lot
of social science principles and methodologies are inspired from similar
counterparts in the natural sciences. A central aspect of our lives
is uncertainty and our struggle to overcome it. Over the years, it
seems that we have found ways to understand the uncertainty in the
natural world by postulating numerous physical laws. 

These physical laws are deductive and are based on three statements
- a specific set of initial conditions, a specific set of final conditions
or explicanda and universally valid generalizations. Combining a set
of generalizations with known initial conditions yields predictions;
combining them with known final conditions yields explanations; and
matching known initial with known final conditions serves as a test
of the generalizations involved. The majority of the predictions in
the physical world hold under a fairly robust set of circumstances
and cannot be influenced by the person making the observation and
they stay unaffected if more people become aware of such a possibility. 

In the social sciences, the situation is exactly the contrary. Popper
{[}2002{]} gives a critique and warns of the dangers of historical
prediction in social systems. In their manifesto, Derman and Wilmott
{[}2009{]}, mention the need to combine art and science in the discipline
of finance. While it is possible to declare that, \textbf{\textit{Art
is Science that we don\textquoteright t know about; and Science is
Art restricted to a set of symbols governed by a growing number of
rules}}, our current state of affairs necessitate that we remain keenly
cognizant of the shortcomings of forecasting. A set of initial conditions
yielding a prediction based on some generalization, ceases to hold,
as soon as many participants become aware of this situation and act
to take advantage of this situation. This means that predictions in
the social sciences are valid only for a limited amount of time and
we cannot be sure about the length of this time, since we need to
constantly factor in the actions of everyone that can potentially
influence a prediction, making it an extremely hard task. 

\textbf{\textit{All attempts at prediction, including both the physical
and the social sciences, are like driving cars with the front windows
blackened out}} and using the rear view mirrors, that give an indication
of what type of path has been encountered and using this information
to forecast, what might be the most likely type of terrain that lies
ahead for us to traverse. The path that has been travelled then becomes
historical data that has been collected through observation and we
make estimates on the future topography based on this. Best results
generally occur, when we combine the data we get in the rear view
mirror with the data we get from the side windows, which is the gauge
of the landscape we are in now, to get a better comprehension of what
lies ahead for us. The quality of the data we gather and what the
past and the present hold then give an indication to what the future
might be. So if the path we have treaded is rocky, then the chances
of it being a bumpy ride ahead are higher. If it has been smooth,
then it will be mostly smooth. Surely, the better our predictions,
the faster we can move; but then again, it is easy to see that the
faster we travel, the more risk we are exposed to, in terms of accidents
happening, if the constitution of the unseen scenery in front of us
shifts drastically and without much warning. 

A paramount peculiarity of the social sciences is that passage on
this avenue is part journey and part race. The roads are muddy, rocky
and more prone to have potholes. This means being early or ahead on
the road brings more winnings. We also have no easy way of knowing
how many people are traveling on this path, either with us, ahead
of us or even after us. As more people travel on the path, it starts
falling apart, making it harder to travel on it, a situation which
is accentuated considering we don't have any vision out front. On
the other hand, let us say, physical science roads, being well paved
and well-constructed using concrete, hold steady for much longer time
durations, so what has been observed in the past can be used to make
durable forecasts that hold for lengthier amounts of time in the future. 

Paich and Sterman {[}1993{]} inquire into decision making in complex
environments and conduct an experiment where subjects must manage
a new product from launch through maturity, and make pricing and capacity
decisions. They demonstrate that decision making in complex dynamic
environments tends to be flawed in specific ways by not accounting
sufficiently for feedback loops, time delays and nonlinearities. Even
with a decent amount of experience, there is no evidence that environments
with high feedback complexity can produce improved decision making
ability. 

The Sweeney and Sweeney {[}1977{]} anecdote about the Capitol Hill
baby-sitting crisis exposits the mechanics of inflation, setting interest
rates and monetary policies required to police the optimal amount
of money. The creation of a monetary crisis in a small simple environment
of good hearted people expounds that even with near ideal conditions,
things can become messy; then in a large labyrinthine atmosphere,
disaster could be brewing without getting noticed and can strike without
much premonition. Taleb {[}2005{]} is an entertaining narrative of
the role of chance in life and in the financial markets. Taleb {[}2010{]}
calls our attention to Black Swan events, which are extremely hard
to detect, highlighting the perils of the prediction business. 

This inability to make consistent predictions in the social sciences
and the lack of an objective measure of value or a True Price Theory
means that is almost impossible for someone to know what a real state
of equilibrium is. Elton, Gruber, Brown and Goetzmann {[}2009{]} review
the concepts related to efficient markets and other aspects of investing;
Kashyap {[}2014{]} explained the pleasures and pitfalls of managing
a portfolio, while emphasizing the cyclical nature of the investment
process. The efficient market hypothesis in spite of being a very
intriguing proposition, can at best claim that markets have a tendency
to move towards being efficient, though a state of equilibrium is
never fully attained since no one has an idea what that state of equilibrium
is and the actions of the participants serves only to displace any
state of equilibrium, if it did exist. The analogy for this would
be a pendulum with perpetual motion; it swings back and forth around
its place of rest with decreasing amplitude and the place of rest
keeps changing with time, starting a new cycle of movement with reinforced
vigour. 

We can then summarize the above with the \textbf{\textit{Uncertainty
Principle of the Social Sciences}}, which can be stated as, \textbf{\textit{\textquotedblleft Any
generalization in the social sciences cannot be both popular and continue
to yield accurate predictions or in other words, the more popular
a particular generalization, the less accurate will be the predictions
it yields\textquotedblright }}. This is because as soon as any generalization
and its set of conditions become common knowledge, the entry of many
participants shifts the equilibrium or the dynamics, such that the
generalization no longer applies to the known set of conditions. 

All our efforts as professionals in the field of financial services,
will then be to study uncertainty and uncover quasi-generalizations;
understand its limitations in terms of what can be the closest states
of pseudo-equilibrium; how long can such a situation exist; what factors
can tip the balance to another state of temporary equilibrium; how
many other participants are aware of this; what is their behaviour
and how is that changing; etc., making our professions a very interesting,
challenging and satisfying career proposition.

\section{Application to Currency Market Making }

With the above discussion in mind, we can turn specifically to how
it applies to market making in financial assets. The increasing use
of algorithms and automation has increased the frequency of trading
for most securities that trade in high volumes. Dempster, M. A. H.,
\& Jones, C. M. {[}2001{]}; Avellaneda and Stoikov {[}2008{]}; Chiu,
Lukman, Modarresi and Velayutham {[}2011{]} and Chlistalla, Speyer,
Kaiser and Mayer {[}2011{]} provide detailed accounts of high frequency
trading and the evolution of various algorithms used towards that
end. The increased frequency of trading means that the bid and offer
quoted for a security also need to be constantly changing. It is common
practice for market makers to set the bid and offer to depend on the
size of the inventory and revise it as the inventory builds up in
either direction. This clearly comes with a number of drawbacks, primary
among which is the lack of change in the quotes due to the rapidly
changing market and the wide variety of variables that capture the
trading conditions. The other participants in this market making system,
which in this case are the counterparties of the market maker, can
observe the quotes and take decisions that will influence the system
and the quoting mechanism may not register these new conditions till
much later. 

Hence to deal with the dynamic nature of the trading and market conditions,
our model has to be adaptive and include a feedback loop that alters
the bid offer adjustment based on the modifications in the market
and trading conditions, without a significant time delay. The market
conditions here refer to factors that are beyond the direct control
of the market maker and this information is usually available publicly
to other participants. Trading conditions refer to factors that can
be influenced by the market maker and are dependent on the trading
book being managed and will be privy only to the market maker and
will be mostly confidential to others. 

The market maker has access to a rich set of trading metrics, which
are not immediately available to other participants. These metrics
can affect the future direction of the price and hence using them
to alter the quote leads to better profits. But given that the trading
conditions are constantly changing, we need to revise the parameters
of the alteration mechanism based on the conditions from the recent
past. This forms a feedback loop that keeps changing the model dynamically
based on what is happening in the market maker\textquoteright s trading
book. As discussed earlier, prediction is a perilous business; hence
it is important to keep the number of parameters to a minimum while
not ignoring any significant causes of change. With this motivation,
we include the changes coming in from different sources by using adequate
yet relatively simple econometrics techniques. This leads to changes
in the model outputs that aid the quotation process and the constant
revision of the model parameters is geared to deal with shifting regimes. 

Any model that automatically updates the quotes is more suited for
instruments that have a high number of transactions within short intervals,
making it hard for traders to manually monitor and adjust the spread;
though this is by no means a stringent requirement. We can use similar
models for illiquid instruments as well and use the quotations provided
by the model as a baseline for further human refinement. We have chosen
currency markets to build the sample model since they are extremely
liquid, Over the Counter (OTC), and hence trading in them is not as
transparent as other financial instruments like equities. Copeland
{[}2008{]} provides a rich discussion on exchange rates and currencies.
The nature of currency trading implies that participants other than
the market marker do not have any idea on the actual volumes traded
and the number of trades. For the purposes of building our model,
we simulate the number of trades and the average size of trades from
a log normal distribution. Norstad {[}1999{]} proves key propositions
regarding normal and log normal distributions. The parameters of the
log normal distributions are chosen such that the total volume in
a certain interval matches the volume publicly mentioned by currency
trading firms. This methodology can be easily extended to other financial
instruments and possibly to any product with an ability to make electronic
price quotations or even be used to periodically perform manual price
updates on products that are traded non-electronically. 

The factors we incorporate in our model to adjust the currency bid-offer
spread are 
\begin{enumerate}
\item The Exchange Rate Volatility 
\item The Trade Count 
\item The Volume 
\end{enumerate}
The exchange rate volatility is publicly observable; and the trade
count and volume, are generally only known to the market maker, in
various instruments over different historical durations in time. The
contributions of each of the factors to the bid-offer adjustment are
computed separately and then consolidated to produce a very adaptive
bid-offer quotation. The subsequent sections consider the calculations
for each factor separately and the consolidation in detail.

\subsection{Exchange Rate Volatility Factor }

This factor is calculated based on the conditional standard deviation
of the exchange rate returns as a function of the lagged conditional
standard deviations and the lagged innovations. 

\[
P_{f}\Longleftrightarrow\sigma_{t}=\alpha\sigma_{t-1}+\beta\varepsilon_{t-1}
\]
\[
P_{f}\text{ is the Price Factor; }\sigma_{t}\text{ is volatility at time }t\text{ ; }\varepsilon_{t-1}\text{ is the innovation at time }t-1\text{ ; }0<\alpha,\beta<1.
\]

Numerous variations to the above formula are possible by extending
it to the GARCH$(p,q)$ type of models. Engle {[}1982{]} is the seminal
work on modeling heteroscedastic variance. Bollerslev {[}1986{]} extends
this technique to a more generalized approach and Bollerslev {[}2008{]}
lists an exhaustive glossary of the various kinds of autoregressive
variance models that have mushroomed over the years. Hamilton {[}1994{]}
and Gujarati {[}1995{]} are classic texts on econometrics methods
and time series analysis that accentuate the need for parsimonious
models. We prefer the simple nature of the sample model, since we
wish to keep the complexity of the system as minimal as possible,
while ensuring that the different sources of variation contribute
to the modification. This becomes important since we are constantly
checking the feedback loop for the system performance. When such a
model is being used empirically, less number of parameters eases the
burden of monitoring; isolating the causes of feedback failure becomes
relatively straight forward; and corrective measures can be quickly
implemented, which could involve tweaking the model parameters. Since
volatility is mean reverting and has a clustering behavior, it is
better to use a model similar to our sample, instead of simply taking
the deviation from a historical average as we use for the other factors
below. A more common variant that is comparable in simplicity to the
one used above is by taking the absolute value of the lagged innovations.
It is left to the practitioner to decide on the exact nature of the
model to use depending on the suitability for their trading needs
and the results they are getting. 

The $t=0$ value of the volatility is calculated based on the standard
deviation of the rate of change of the exchange rates from a historical
period. We use a 30 day historical period to calculate the initial
volatility. 

We model the innovation, $\varepsilon$, as the rate of the change
of the exchange rates with respect to time. This is calculated as
the natural logarithm of the ratio of the exchange rates at two consecutive
time periods. In the sample designed to demonstrate the model, we
use the time interval between consecutive rates to be 60 seconds.
\[
\varepsilon_{t-1}=\ln\left(\frac{R_{t-1}}{R_{t-2}}\right)
\]
\[
\varepsilon_{t-1}\text{ is the innovation at time }t-1\text{ ; }R_{t-1}\text{ is the exchange at t-1}.
\]

\subsection{Trade Count Factor }

We first calculate the historical average of the trade count during
a certain time interval. In the sample model, the historical average
is based on a 30 day rolling window. The time interval is 60 seconds.
We measure how the trade count for the latest time interval differs
from the historical average. This is measured as the natural logarithm
of the ratio of the trade count for the latest time interval to the
historical average of the trade count.
\[
TC_{f}\Longleftrightarrow\ln\left(\frac{TC_{i}}{TC_{avg}}\right)*\gamma
\]

$TC_{f}$ is the Trade Count Factor; $TC_{i}$ is the Trade Count
during minute $i$ or during a certain interval of consideration; 

$TC_{avg}=\left(\text{Number of Trades in a Month}\right)/\left(\text{Number of Trading Days in the Month}*\text{Number of Minutes in a Day}\right)$.
It is calculated as a rolling average; $\gamma$ is the parameter
that is used to scale the trade count factor into a similar size as
the price factor. It is the average of the price factor over a suitable
historical range. We use the average over the last thirty days. 

Note: A 30 day rolling window results in the historical averages getting
updated every trading day.

\subsection{Volume Factor }

We first calculate the historical average of the volume during a certain
time interval. In the sample model, the historical average is based
on a 30 day rolling window. The time interval is 60 seconds. We measure
how the volume for the latest time interval differs from the historical
average. This is measured as the natural logarithm of the ratio of
the volume for the latest time interval to the historical average
of the volume. 
\[
V_{f}\Longleftrightarrow\ln\left(\frac{V_{i}}{V_{avg}}\right)*\gamma
\]

$V_{f}$ is the Volume Factor; $V_{i}$ is the Volume in USD during
minute $i$ or during a certain interval of consideration; $V_{avg}=\left(\text{Volume in a Month}\right)/\left(\text{Number of Trading Days in the Month}*\text{Number of Minutes in a Day}\right)$.
It is calculated as a rolling average; $\gamma$ is the parameter
that is used to scale the volume factor into a similar size as the
price factor. It is the average of the price factor over a suitable
historical range. We use the average over the last thirty days. 

Note: A 30 day rolling window results in the historical averages getting
updated every trading day.

\section{Consolidation of the Three Factors }

The three factors are consolidated by using a weighted sum. In the
sample model, all three factors are equally weighted. Henceforth,
the consolidated factor will be referred to as the spread factor.
Where required, depending on the financial instrument, each of the
three individual factors can be scaled down to be in the order of
the magnitude of the adjustment we want to make to the bid and the
offer. We do not require this step for our sample model, since the
order of magnitude of the spread factor is in the same region as the
adjustment to the spread we wish to make. We also calculate the historical
average and standard deviation of the spread factor. In the sample
model, the historical average and standard deviation are based on
a 30 day rolling window. We consider the spread factor to be a normal
distribution with mean and standard deviation equal to the 30 day
historical average and standard deviation. When the spread factor
is more than a certain number of standard deviations to the right
of the historical average of the spread factor, we increase the bid-offer
spread. If the spread factor is more than a certain number of standard
deviations to the left of the historical average, we decrease the
bid-offer spread. In the sample model, we consider half a standard
deviation to the right and a third of a standard deviation to the
left of the mean. The increase or decrease of the bid-offer spread
is proportional to the magnitude of the spread factor. The maximum
spread change is limited to an appropriate pre-set threshold for both
the upper and lower limit.
\[
S_{rf}\Longleftrightarrow w_{p}*P_{f}+w_{tc}*TC_{f}+w_{v}*V_{f}
\]
\begin{eqnarray*}
S_{f}:\left\{ \left.S_{rf}\right|\text{if }\left[S_{rf}\leq\left(\mu_{S_{rf}}+\frac{\sigma_{S_{rf}}}{m}\right)\right]\text{ then}\right.\\
\text{if }\left[S_{rf}<\left(\mu_{S_{rf}}-\frac{\sigma_{S_{rf}}}{n}\right)\right]\text{ then}\\
\left[\mu_{S_{rf}}-\frac{\sigma_{S_{rf}}}{n}\right]\\
\text{else}\\
\left[S_{rf}\right]\\
\text{end if}\\
\text{else}\\
\left[\mu_{S_{rf}}+\frac{\sigma_{S_{rf}}}{m}\right]\\
\left.\text{end if}\vphantom{\left.S_{rf}\right|\text{if }\left[S_{rf}\leq\left(\mu_{S_{rf}}+\frac{\sigma_{S_{rf}}}{m}\right)\right]\text{ then}}\right\} 
\end{eqnarray*}

$S_{rf}$ is the raw Spread Factor; $\mu_{S_{rf}}$ is the rolling
average of the raw Spread factor; $\sigma_{S_{rf}}$ is the rolling
standard deviation of the raw Spread Factor; $S_{f}$ is the spread
factor after adjusting for the upper and lower bounds; $m,n\in\mathcal{R}$;
we have set $m=2$ and $n=3$; $w_{p}$ is the weight for the Price
Factor; $P_{f}$ is the Price Factor; $w_{tc}$ is the weight for
the Trade Count Factor; $TC_{f}$ is the Trade Count Factor; $w_{v}$
is the weight for the Volume Factor; $V_{f}$ is the Volume Factor.

Note: A 30 day rolling window results in the historical average and
standard deviation getting updated every trading day.

\section{Dataset Construction }

To construct a sample model, we need the following data items: the
price, the trade count and the volume of the security over different
time intervals. We have chosen the currency markets since it is an
ideal candidate for a dynamic quotation model, but the price is not
publicly disclosed as in the equity markets. We take the average of
the high, low, open and close prices over a certain interval as a
proxy for the trade price. Many market making firms disclose such
a data set at different intervals facilitating the creation of a reasonable
hypothetical price. The data is available over our chosen interval
of one minute as well. 

The trade count and trade volume over a minute are not publicly available.
But many providers disclose total quarterly, total monthly and average
daily volumes. The volume over a minute is the product of the number
of trades and the size of each trade during that minute. We can pick
random samples from a log normal distribution to get the trade count
and trade size for each minute. The mean and standard deviation of
the log normal distributions can be set such that the total volume
will match the publicly disclosed figure. We can make an assumption
that there will be sixty trades on average in a minute and set the
average trade size based on the total volume. Please see endnote {[}1{]}
and {[}2{]} in the references for further details on the publicly
available data sets and Appendix \ref{sub:Model-Parameters-and} for
details of the model parameters. Any market maker wishing to use this
model can easily substitute the simulated variables with the actual
values they observe.

\section{Model Testing Results }
\begin{enumerate}
\item The model was tested on a time horizon between 24-Jul-2013 to 24-Oct-2013.
The currency pair used was the EUR-USD currency pair and the hypothetical
trade price is the average of the high, low, open and close during
a certain interval, which in our case was a minute. The high, low,
open and close is publicly available from a number of providers. 
\item The ideal starting historical values are to be calculated based on
data from the month preceding this period. Other shorter time intervals
can be considered as appropriate to the needs of the specific trading
desks. 
\item The P\&L increase for this time period was USD \$513,050. P\&L breakdown
by trading day and by trading hour are included in Appendix \ref{sub:USD-Profit-=000026},
\ref{sub:USD-Profit-=000026-1}, \ref{sub:Key-Metrics-by} and \ref{sub:Key-Metrics-by-1}.
It is important to keep in mind that most liquid currencies trade
continuously from Monday morning Asia time to Friday evening US time. 
\item The spread was increased 47,347 times; decreased 48,244 times; the
spread factor was greater than the upper bound on 19,605 times and
lower than the lower bound on 27,535 times. 
\item The volume that was affected by the increased spread was approximately
444.95 Billion; volume affected by the decreased spread was 443.19
Billion. 
\item More detailed results are included in the Appendices.
\end{enumerate}

\section{Improvements to the Model }
\begin{enumerate}
\item We can skew the change in the bid offer spread to be more on the bid
or the offer side based on the buy and sell volumes. We have not considered
this exclusively in our model since we only look at the change in
the spread and not on which side of the quote the change happens.
It is simple to adjust both sides equally or be cleverer in how we
split the total spread change into the bid or the offer side. 
\item The assumption of normality and the use of a log normal distribution
can be relaxed in favor of other distributions. It is also possible
to use different distributions that change over time, as a result
of the feedback we receive from the system. This is a more realistic
portrayal of empirical data which tend to fall into different distributions
as regimes change. 
\item Each of the variables can be modeled using more advanced econometric
techniques like the GARCH$(p,q)$ model. Care needs to be taken that
the additional parameters do not impact the feedback loop and when
results are not satisfactory, we can easily investigate the reason
for issues. 
\item For simplicity, we have ignored the question of negative spreads or
reverse quotes, where the bid is greater than the offer, resulting
in a crossed market. This can happen when the magnitude of the spread
factor is greater than the difference between the bid and the offer.
This can be handled easily by reducing the size of the spread factor
when such an event occurs. Additional ways to handle this are considered
in the below points. 
\item The model can be made to adapt its scaling factors, the alpha and
the beta so that the difference in the average of the increase and
the average of the decrease in the spread are equal over a certain
time period. With this, the overall spread change stays the same and
the market maker is seen to be quoting competitive spreads, though
this results in better profitability based on the volume and price
movements it is experiencing. See Appendix-I for details on the model
parameters. 
\item In our current model, we limit the size of the spread change on both
the positive and negative sides depending on the value of the spread
factor. A variation to this can be to change the spread only when
the spread factor lies above or below a certain threshold. The spread
change can be a constant value; or two constant values, one for the
increment and one for the decrement or it can be made to depend on
the spread factor as well. 
\item The consolidated spread factor computed as the weighted sum of the
exchange rate volatility, trade count and volume factors can be made
to depend more on the volatility and trade count by adjusting the
corresponding weights. 
\item The time interval considered for the factors is 60 seconds. Smaller
time intervals will result in better performance for currency markets.
Larger time intervals might be more suited for other securities. 
\item The rolling average can be taken over shorter or longer intervals
depending on the results and the security under consideration. It
is also possible to weight different contributors to the average differently
resulting in a Moving Average model. 
\item The trade count and volume factors can also be modeled similar to
the Exchange Rate Volatility Factor. The point to bear in mind is
that the exchange rate volatility is mean reverting and the trade
count and volume factors have always had an upward trend. This is
because we expect more trading to happen and all trading desks are
bullish about their activities. Given the volume projections, we can
expect the upward trajectory for these two factors to continue. For
30 day rolling windows, we can assume that the trade count and volume
follow a mean reverting property. For our purpose, the deviations
from the 30 day historical average for the trade count and volume
factors produce satisfactory results. 
\item A central question is whether the changing spread will have a negative
impact on the volumes traded and hence on the overall profitability
of the desk. This needs to be monitored closely and the size of the
changes need to be adjusted accordingly. 
\item Other factors can be included, like the percentage of flow handled
by the market marker to the average flow in that currency pair over
the course of a trading day. This factor indicates the extent of monopoly
that the market maker enjoys and indicates pricing power. This ratio
can be used to adjust the spread in the favor of the market maker
or in the feedback loop to tweak the parameters that are used for
other factors.
\end{enumerate}

\section{Conclusion }

The need for a dynamic quotation model comes from the feature of the
social sciences and trading, where observations coupled with decision
making can impact the system. This aspect was illustrated in detail
and summarized as the uncertainty principle of the social sciences.
To deal with this phenomenon, we need a feedback mechanism, which
incorporates trading conditions into the quotation process, without
too much of a temporal lag. 

A model was constructed, using price, trade count and volume factors
over one minute intervals, to vary the quotes being made. The models
constructed are rich enough to capture the effect of the various relevant
factors, yet simple enough to accord constant monitoring and to ensure
the effectiveness of the feedback loop. The real test of any financial
model or trading strategy is the effect on the bottom line and hence
when we looked at the performance of our methodology, we found the
positive effect on the P\&L to be significant, without too much of
a change to the way the trading happens or an accompanying increasing
in risk or leverage of the trading desk. 

Numerous improvements to the model are possible and can be considered
depending on the type of instrument being traded and the technology
infrastructure available for trading. Future iterations of this study
will look to extend this methodology to other asset classes.

\section{References and Notes }
\begin{enumerate}
\item For further details on the publicly available datasets, see \href{http://forexmagnates.com/fxcm-posts-records-quarterly-revenues-and-july-volume-metrics/ and http://ir.fxcm.com/releasedetail.cfm?ReleaseID=797967}{http://forexmagnates.com/fxcm-posts-records-quarterly-revenues-and-july-volume-metrics/ and http://ir.fxcm.com/releasedetail.cfm?ReleaseID=797967}
\item The author has utilized similar algorithms for market making in various
OTC as well as exchange traded instruments for more than the last
ten years. As compared to the sample model, the interval trade count
and volumes used in the empirical model were the actual observations;
yet the overall results are somewhat similar.
\item Avellaneda, M., \& Stoikov, S. (2008). High-frequency trading in a
limit order book. Quantitative Finance, 8(3), 217-224.
\item Bodie, Z., Kane, A., \& Marcus, A. J. (2002). Investments. International
Edition.
\item Bollerslev, T. (1986). Generalized autoregressive conditional heteroskedasticity.
Journal of econometrics, 31(3), 307-327.
\item --. (2008). Glossary to arch (garch). CREATES Research Paper, 49.
\item Chiu, J., Lukman, D., Modarresi, K., \& Velayutham, A. (2011). High-frequency
trading. Standford, California, US: Stanford University.
\item Chlistalla, M., Speyer, B., Kaiser, S., \& Mayer, T. (2011). High-frequency
trading. Deutsche Bank Research, 1-19.
\item Copeland, L. S. (2008). Exchange rates and international finance.
Pearson Education.
\item Dempster, M. A. H., \& Jones, C. M. (2001). A real-time adaptive trading
system using genetic programming. Quantitative Finance, 1(4), 397-413.
\item Derman, E., \& Wilmott, P. (2009). The Financial Modelers' Manifesto.
In SSRN: http://ssrn. com/abstract (Vol. 1324878).
\item Elton, E. J., Gruber, M. J., Brown, S. J., \& Goetzmann, W. N. (2009).
Modern portfolio theory and investment analysis. John Wiley \& Sons.
\item Engle, R. F. (1982). Autoregressive conditional heteroscedasticity
with estimates of the variance of United Kingdom inflation. Econometrica:
Journal of the Econometric Society, 987-1007.
\item Gujarati, D. N. (1995). Basic econometrics, 3rd. International Edition.
\item Hamilton, J. D. (1994). Time series analysis (Vol. 2). Princeton university
press.
\item Hull, J. C. (1999). Options, futures, and other derivatives. Pearson
Education India.
\item Kashyap, R. \textquotedbl{}The Circle of Investment.\textquotedbl{}
International Journal of Economics and Finance, Vol. 6, No. 5 (2014),
pp. 244-263. 
\item Marcus, A. J., Bodie, Z., \& Kane, A. (2002). Investments. McGraw
Hill.
\item Norstad, J. (1999). The normal and lognormal distributions.
\item Paich, M., \& Sterman, J. D. (1993). Boom, bust, and failures to learn
in experimental markets. Management Science, 39(12), 1439-1458. 
\item Popper, K. R. (2002). The poverty of historicism. Psychology Press.
\item Sweeney, J., \& Sweeney, R. J. (1977). Monetary theory and the great
Capitol Hill Baby Sitting Co-op crisis: comment. Journal of Money,
Credit and Banking, 86-89.
\item Taleb, N. (2005). Fooled by randomness: The hidden role of chance
in life and in the markets. Random House LLC.
\item --. (2010). The Black Swan: The Impact of the Highly Improbable Fragility.
\item Tuckman, B. (1995). Fixed Income Securities-Tools for Today\textquoteright s
Markets.
\end{enumerate}

\section{Appendix}

\subsection{\label{sub:Model-Parameters-and}Model Parameters and Key Metrics }

\begin{figure}[H]
\includegraphics[width=15cm,height=20cm,keepaspectratio]{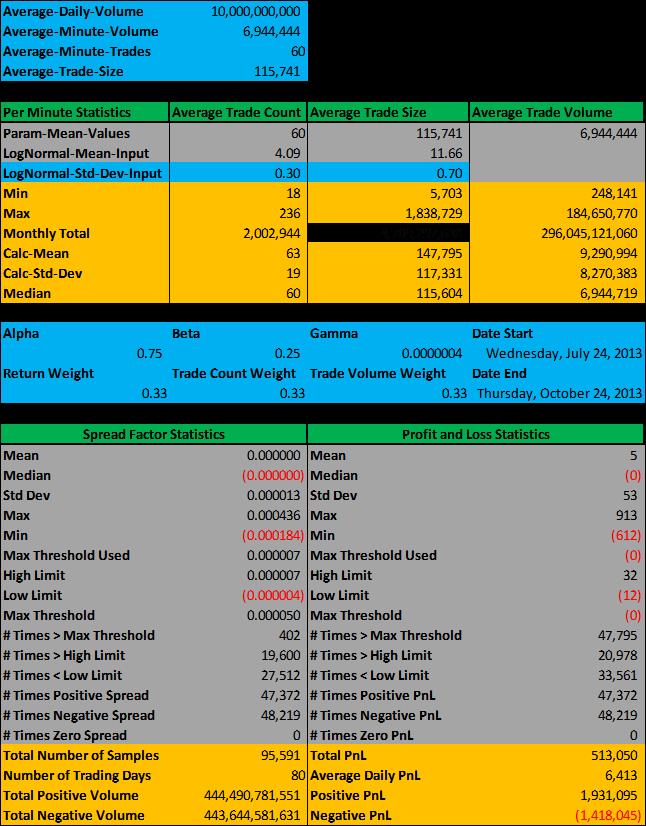}

\caption{Model Parameters and Key Metrics\label{fig:Model-Parameters-and}}

\end{figure}

In Figure \ref{fig:Model-Parameters-and}, values in blue are model
parameters that can be used to optimize the model. In the sample model,
these act as user inputs and can be changed to see how the model behaves
under different conditions. Values in green are common categories
that apply to different metrics. Values in bright yellow are important
metrics, some of which form a key part of the feedback loop and it
would be good to monitor these closely. Alpha and Beta are the parameters
used to model the volatility of the price. Gamma is the parameter
that is used to scale the trade and volume factors into a similar
size as the price factor.

\subsection{\label{sub:USD-Profit-=000026}USD Profit \& Loss by Trading Day}

\begin{figure}[H]
\includegraphics[width=15cm,height=20cm,keepaspectratio]{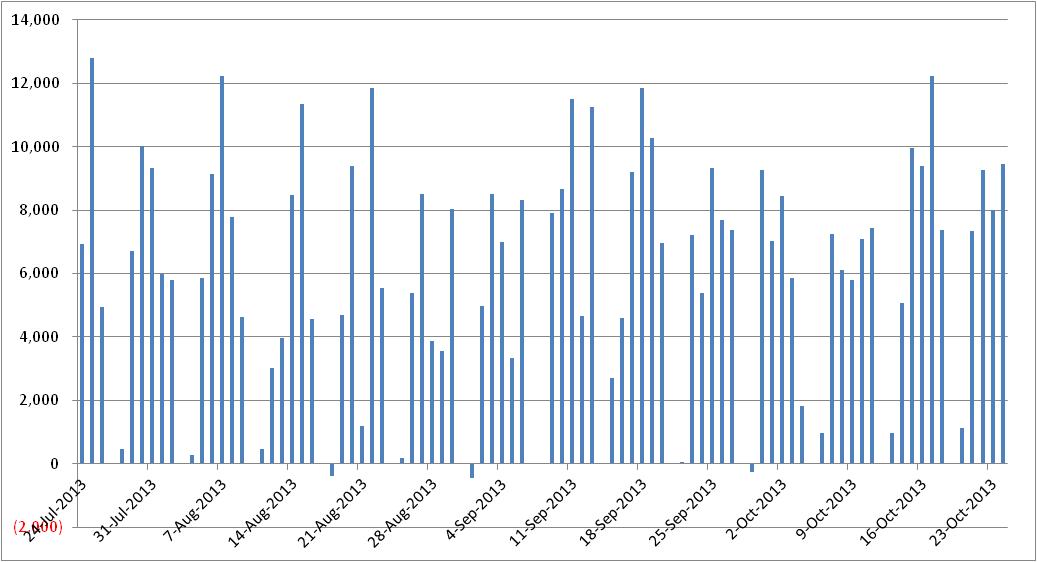}

\caption{USD Profit \& Loss by Trading Day}

\end{figure}

\subsection{\label{sub:USD-Profit-=000026-1}USD Profit \& Loss by Trading Hour}

\begin{figure}[H]
\includegraphics[width=15cm,height=20cm,keepaspectratio]{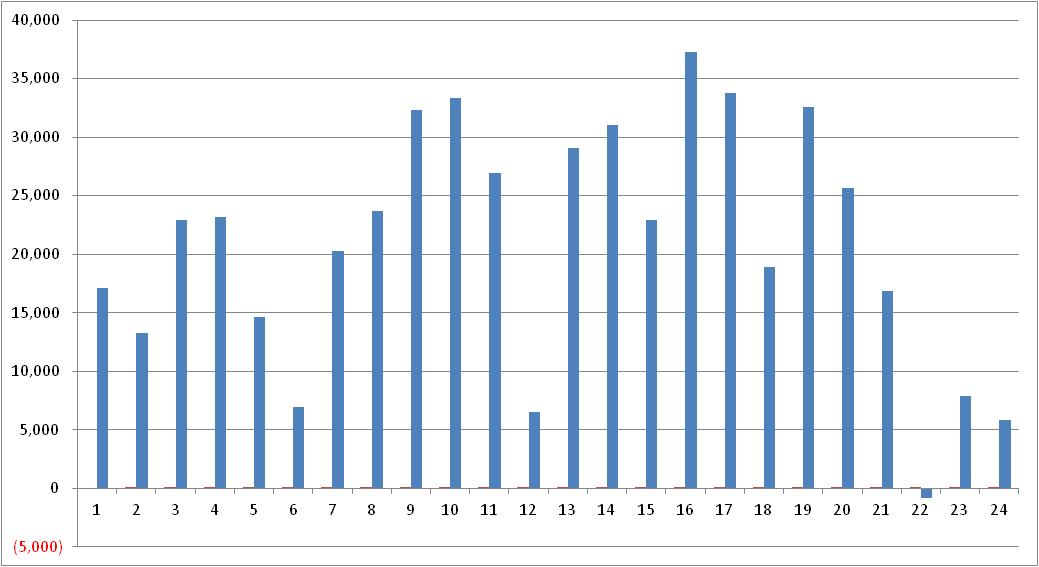}

\caption{USD Profit \& Loss by Trading Hour}

\end{figure}

\subsection{\label{sub:Key-Metrics-by}Key Metrics by Trading Day}

\begin{figure}[H]
\includegraphics[width=16cm,height=22cm]{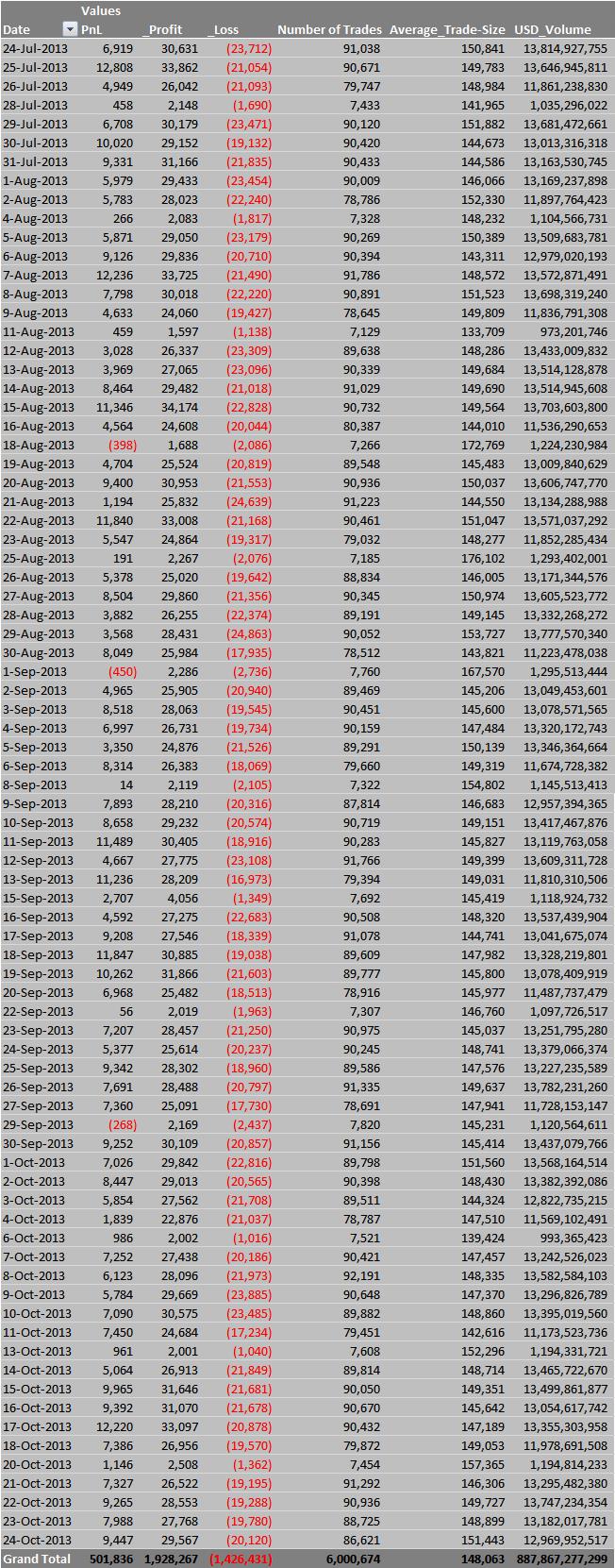}

\caption{Key Metrics by Trading Day}

\end{figure}

\subsection{\label{sub:Key-Metrics-by-1}Key Metrics by Trading Hour}

\begin{figure}[H]
\includegraphics[width=15cm,height=20cm,keepaspectratio]{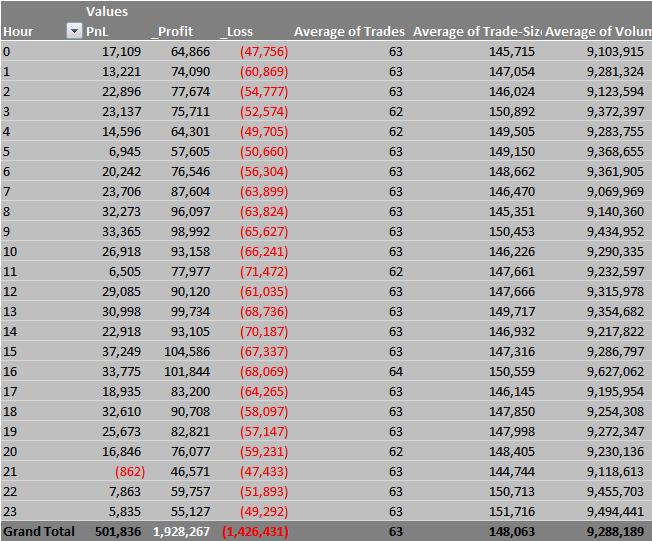}

\caption{Key Metrics by Trading Hour}

\end{figure}

\end{document}